# PERFORMANCE EVALUATION OF ADVANCED CONGESTION CONTROL MECHANISMS FOR COAP

By

CHANDRA SEKHAR SANABOINA * TEJESWAR ELURI **

* Assistant Professor, Department of Computer Science and Engineering, JNTUK, Kakinada, Andhra Pradesh, India.
** M.Tech Student, JNTUK, Kakinada, Andhra Pradesh, India.

*ABSTRACT*

*In Internet of Things (IoT), the simple IPv6 capable electronic devices with limited hardware resources like memory and power resources are called constrained devices. Congestion is a major issue in network communications of these devices. To solve congestion in networks of constrained devices, Internet Engineering Task Force (IETF) had designed Constrained Application Protocol (CoAP). CoAP deals congestion with a basic Congestion Control (CC) mechanism called Default CoAP. Afterward, CoCoA and CoCoA+, an Internet-draft-recommendations, has been introduced as the elective CC mechanisms for CoAP. However, limited evaluations had done on these CC mechanisms of CoAP. In this paper, the performance evaluation of Default CoAP, CoCoA, and CoCoA+ Congestion control mechanisms are evaluated through Simulations in different network topologies and varied in different Link Delivery Ratios (LDR) of sensor nodes in constant traffic scenario by using Cooja Simulator. The simulation results are generated and CoCoA+ shows a better performance in constant traffic scenario comparing to Default CoAP and CoCoA.*

## INTRODUCTION

Internet of things (IoT) is a new age of innovation that each object regardless of devices or human could be associated with the Internet. There are numerous sorts of wireless protocols (like IEEE 802.11 Series, 802.15 Series, ZigBee, and so on.) for network communication between devices. In any case, a lot of little devices cannot associate with the internet due to their constrained resources.

Internet Engineering Task Force (IETF) has designed a lightweight Protocol called Constrained Application Protocol (CoAP) by the IETF open standard RFC 7252 (Shelby et al., 2014). CoAP runs on UDP/IPv6 by default and it can likewise be executed over different channels like TCP, DTLS, or SMS. CoAP depends on the request/response type communication model. It was intended for machine-to-machine (M2M) communication yet was adjusted in IoT also, with help on gateways, high-level servers, and venture integration. CoAP is a Representational State Transfer Style (RESTful) protocol that offers the activities, such as GET, PUT, POST, and DELETE to control resources on servers. CoAP is proposed to be a lightweight contrasting option to HTTP. It is designed as the main Application layer protocol to be utilized by IoT gadgets for IP-based HTTP-like cooperations.

In IoT, network Communications in constrained devices congestion is a major problem. Network congestion happens when the produced traffic load reaches the network limit. Traffic loads that can cause congestion are probably going to occur in CoAP communications when messages between substantial quantities of devices are traded.

Congestion Control mechanisms have the ability to detect congestion in the network and apply measures to avoid congestion for reliable network performance. Investigation of CC mechanisms for constrained devices that runs CoAP is also important. However, performance evaluations have to confirm which congestion control mechanism is suitable for CoAP. In this paper, the Cooja simulation environment is used, which is part of the Contiki Operating system (OS) toolset, to simulate CoAP CC mechanisms.

## 1. Related Work

Congestion Control/Advanced (CoCoA) (Bormann et al., 2014) are assessed with three different network topologies, in particular, a grid topology, a dumbbell topology, and a chain topology with various NSTART parameter values. The outcomes demonstrate that CoCoA has preferred execution in throughput over the Default CoAP. CoCoA can reduce MAC buffer overflows (Betzler et al., 2013).

By analyzing default CoAP and CoCoA, a few downsides of their execution are distinguished. By alterations of these CC mechanisms with new generation Variable Backoff Factor (VBF) and a new RTO aging mechanism (Betzler et al., 2015) introduced a new CC mechanism for CoAP: Congestion Control Advanced/Plus (CoCoA+). The execution aftereffects of default CoAP, CoCoA, and CoCoA+ demonstrate that CoCoA+ can outperform default CoAP and CoCoA regarding PDR upgrades of up to 19.8% and a decrease of end-to-end delays amid bursts of packets of up to 31.2% are estimated in contrast with default CoAP and CoCoA.

## 2. CoAP Congestion Control Mechanism

CoAP is a Representational State Transfer Style (REST-ful) protocol developed for constrained devices which permits connection among clients and servers over the internet through UDP/IPv6 (Fielding & Taylor, 2000). CoAP plays a major role in the environment of constrained devices. CoAP can be utilized for the control of resources, for example, sensor estimations or changing actuator states. CoAP works over UDP. It characterizes four kinds of messages:

- *Confirmable (CON):* The confirmable messages require an acknowledgment response from the receiver in order to provide a reliability functionality.
- *Non-confirmable (NON):* The non-confirmable message does not require an acknowledgment from the receiver.
- *Acknowledgment (ACK):* Acknowledges the confirmable messages.
- *Reset (RST):* The reset is used instead of ACK in case that CON or NON cannot be processed.

Since UDP does not execute end-to-end reliability but it might be rather required by the application. When an application requires end-to-end reliability, (CON) messages are utilized for the (ACK) reply from destination endpoint. If not (NON) messages are utilized. In default CoAP, when a confirmable message is sent, CoAP haphazardly picks a Retransmission timeout (RTO) value from the interim between [2 s, 3 s] for the message retransmission. When timer of the RTO values terminates and source of the message transmission has not gotten an ACK from the destination endpoint, a message loss will occur and the CoAP message will be retransmitted (Shelby et al., 2014).

Sometimes after retransmission timeout, both client CON message and server ACK message collides with each other and creates congestion. To avoid congestion in the network, a Binary Exponential Backoff (BEB) is applied, multiplying the RTO estimation of the retransmitted packet (Balandina et al., 2013). The main contrast in the conduct of CoCoA contrasted with Default CoAP is the utilization of Round-Trip Time (RTT) estimations to ascertain the RTO of the primary transmission of a CoAP message. CoCoA runs two RTO estimators in parallel for every destination endpoint as a strong and weak RTO estimator that is refreshed when estimating strong RTTs or weak RTTs. The process of RTT and RTO calculations can be seen in below Figure 1.

The measurement of an RTTX is used to refresh RTOX as

$$RTO_x = RTT_x + K_x * RTTVAR_x$$

where X means strong or weak.

$RTT_{strong}$ is computed when an ACK is gotten after the primary transmission of a CON message. $RTT_{weak}$ is computed when an ACK is gotten after primary Re-transmission of a CON message.

As per the CoCoA, when $RTO_{strong}$ or $RTO_{weak}$ is refreshed subsequent to getting an RTT estimation, a general RTO ($RTO_{overall}$) is recalculated.

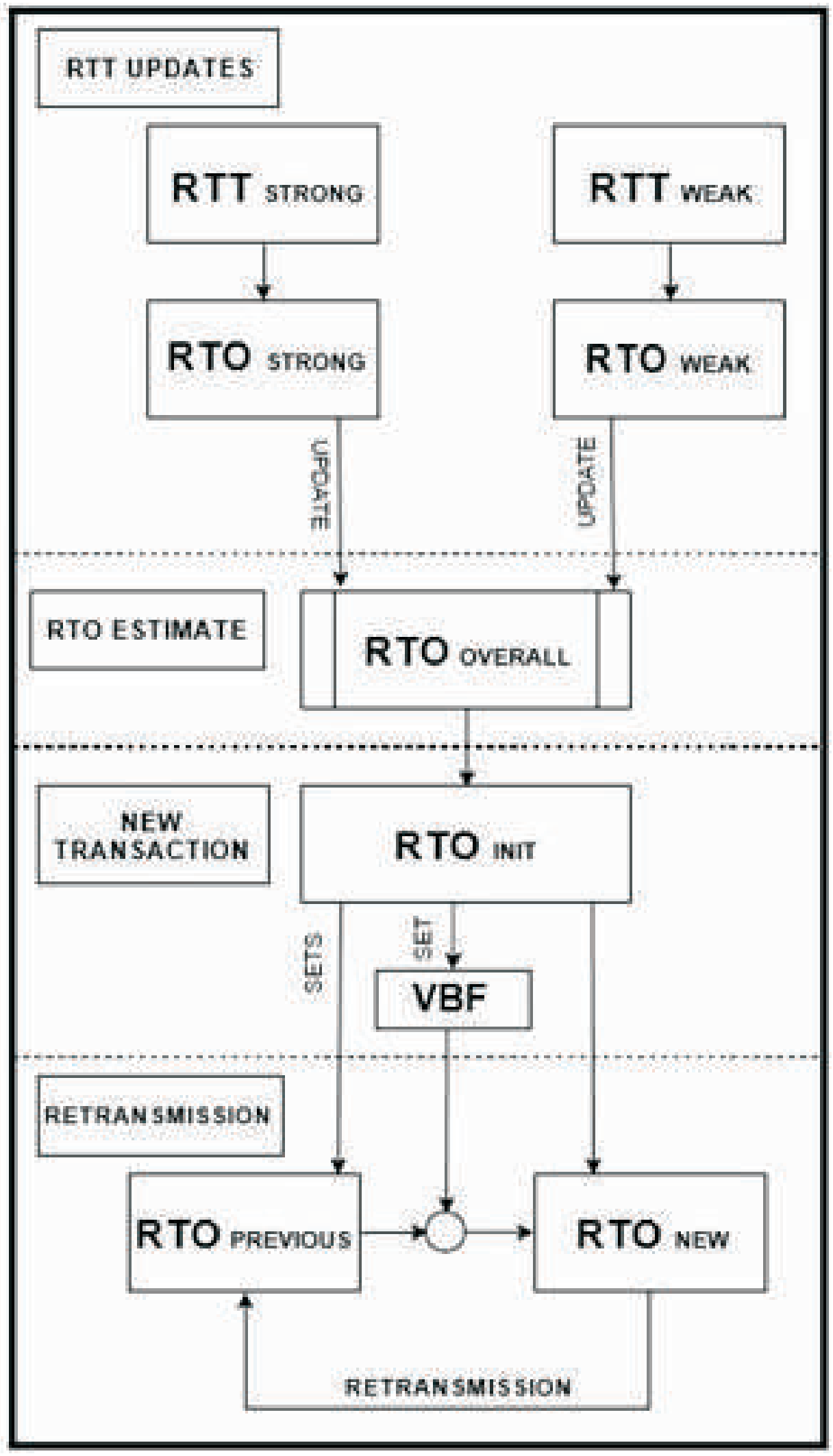

Figure 1. Review of various RTO Factors used to Maintain and Refresh the RTO State Data

$$RTO_{Overall} = 0.5 * RTO_X + 0.5 * RTO_{Overall}$$

In case of CoCoA+, a swap of the BEB utilized for retransmissions by a Variable Back off Factor (VBF) is utilized. The VBF is a vital change to the back off conduct utilized by default CoAP and CoCoA. Rather than multiplying the past RTO value ($RTO_{previous}$) to get the RTO applied to the following retransmission ($RTO_{new}$), it is multiplied by a variable.

$$RTO_{(new)} = RTO_{(previous)} * VBF$$

where VBF relies upon the $RTO_{init}$ of a CoAP trade as below:

$$VBF_{(RTOinit)} = \left\{ \begin{array}{ll} 3, & RTO_{init} < 1s \\ 2, & 1 \leq RTO_{init} \leq 3s \\ 1.3, & RTO_{init} > 3s \end{array} \right\}$$

CoCoA does not have a Decay mechanism for long RTOs that may wind up out of date after longer times of not acquiring fresh RTT estimations. CoCoA+ had presented another RTO mechanism which expresses that if $RTO_{overall}$ is bigger than the base RTO from the default CoAP particular of 2 s, and it is not refreshed amid in excess of 30 s, on the following transmission the $RTO_{overall}$ is refreshed as:

$$RTO_{Overall} = (2 + RTO_{Overall})/2s$$

The above equation revives the $RTO_{overall}$ of an estimator to a value that is nearer to the default estimation of 2 s (Betzler et al., 2015).

## 3. Evaluation Setup

In this section, points of interest on the simulation setup was used to do performance evaluations of the three CC mechanisms. This incorporates the configuration of the simulator, constant traffic scenario, the network topologies, and the performance metric used to do the performance assessments.

### *3.1 Contiki Operating System*

Contiki is a high level multitasking operating system developed for wireless sensor networks and constrained devices (Dunkels et al., 2004). Contiki OS allows small devices with limited hardware resources like RAM, memory, and network capacities to connect to the internet using protocols (IPv4 and IPv6) with low power consumption.

Contiki has a built-in TCP/IP protocol stack. Contiki OS configurations are as small as 2 kilobytes of RAM and 20 kilobytes of ROM. Various communication protocol stacks were used by Contiki such as Rime and uIP/uIP6. These communication stacks are lightweight and designed for Low Power radio sensors. Rime has many communication features like LAN Broadcasting and reliable bulk data transfer. These communication stacks were widely used by companies in systems like satellites.

### *3.2 Cooja Simulator*

Cooja is a Wireless Sensor Network (WSN) simulator which is part of Contiki OS toolset. It is an adaptable Java-based simulator which bolsters utilizing C programs to create application software by Java Native Interface. Cooja main feature is the assessment of the remote sensor nodes equipment that considers the equipment determinations and handling capacities of real nodes (Osterlind et al., 2006).

In Cooja, the binary image of a node can be transferred into the virtual nodes, where the accumulated program code is then executed with the assessment model of the chosen node composed amid the simulations as though it were a real node. Cooja's another feature is that application developer can modify the parts of the simulation environment without changing any Cooja fundamental code.

It implies that the framework can be added to new parts, for example, interfaces, modules, and radio mediums. With these points of interest of Cooja, usage of simulation are done with various conditions and framework settings, for example, packet generation rates, diverse MAC protocols, and distinctive network topologies.

*3.3 Simulations*

Simulations does not suffer from the effects of the firewall, connection issues of internet protocol (IPv6/IPv4), interfaces, and commotion. Simulations give a perfect model of working which endeavors to simulation protocols as nearly as conceivable to their suggestions and formats, to comprehend their conduct and structure. Distinctive topologies and settings can be experimented.

*3.3.1 Need for Simulations*

A simulator is an exceptionally helpful tool for the application programming advancement in WSNs. In past, the code improvement was exceptionally troublesome for clients because of the long compilation and troubleshoots time and the transplant issue of the program. In the case of the simulator, clients can acquire numerous advantages amid the product improvement stage and vast scale test stage. In this paper, CC mechanisms will be evaluated in constant traffic scenario.

*3.4 Constant Traffic Scenario*

The performance metric used to evaluate the simulation results is the overall PDR.

- *Overall PDR:* the aggregate sum of received CoAP messages over the amount of sent CoAP messages in a given time interim. PDR is a marker of how reliable the network is and whether losses are expected.

CoAP message losses may prompt network breaking down at the application. Amid a congestion state, the probability of losses is high. A successful CC mechanism ought to have the capacity to recognize the condition of congestion and take measures to solve it. As a result of applying these measures, CC mechanisms expand the condition of CoAP against message losses, in this way conveying higher PDR values.

From the overall PDR, the network throughput in terms of the carried load against offered load can be determined. Figure 2 demonstrates the default communication protocol stack created by IETF along with implementation in Contiki which is used to evaluate in this paper.

Contiki is built-in with the Erbium CoAP implementation, with observe (Hartke, 2015) and Blockwise transfers (Bormann & Shelby, 2016).

- Erbium is a REST Engine and CoAP Implementation for Contiki. UDP over IPv6 networking is chosen. IPv6 over Low-power personal area network (6lowpan) is chosen for creating a mesh network over the nodes (Kovatsch et al., 2011).
- Radio Duty Cycling (RDC) is used for making nodes in standby mode and Contiki Mac Retransmissions are turned ON so that data packets can be retransmitted (Youssef et al., 2014).
- In the Physical (PHY) and MAC layers, the nodes implement IEEE802.15.4, using a data transmission rate of 250 kbps in the 2.4 GHz radio band. TMote Sky nodes from Moteiv corporation, which supports this configuration of the ContikiOS stack are used in the simulations.
- MSP430-4.7.0 GCC compiler is used for compilation as Contiki 3.0 does not support MCU- MSP430F1611. Therefore, MSP430-4.7.0, experimental GCC compiler which gives the sensor nodes a 20-bit extra memory is used for the simulating the three congestion control algorithms.

The evaluations of Default CoAP, CoCoA, and CoCoA+

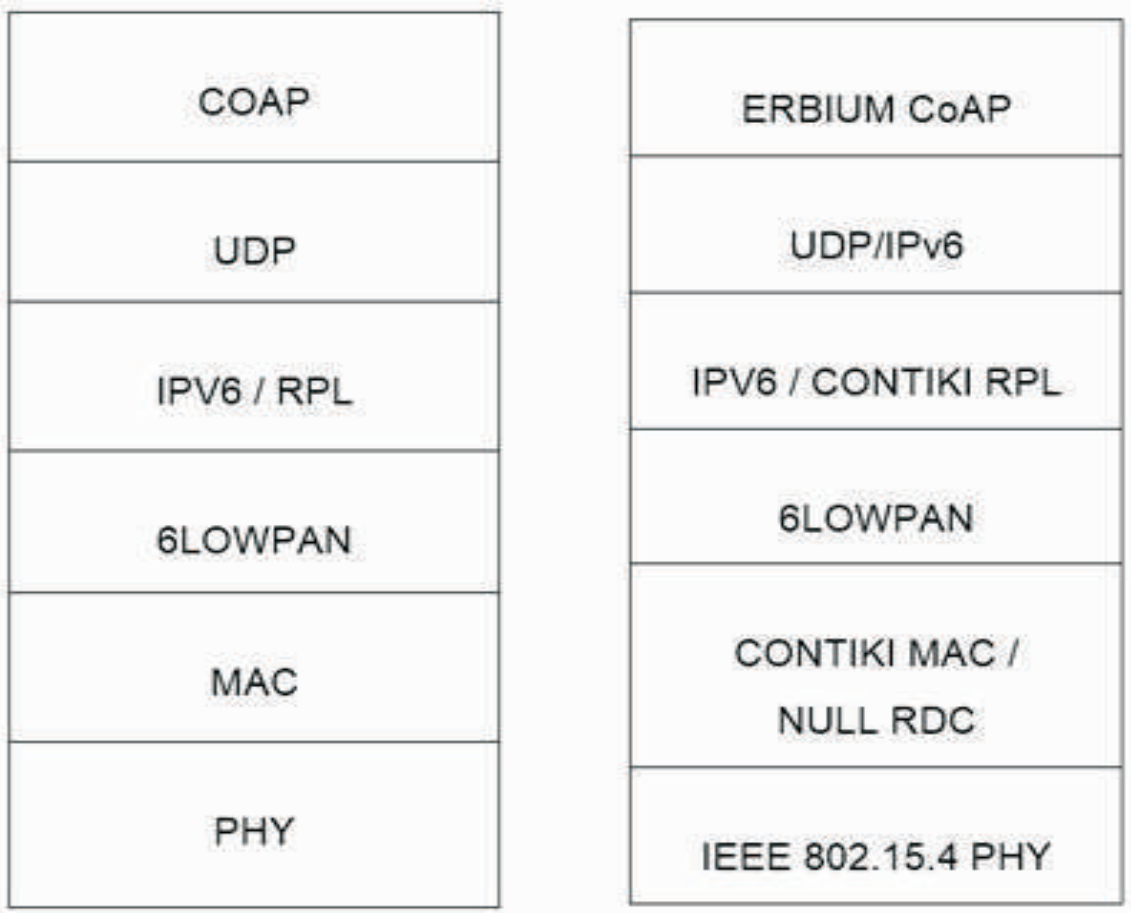

Figure 2. Examination of the IETF Communication Protocol Stack (Left) and its Implementation in Contiki (Right)

CC mechanism will be analyzed in constant traffic scenario. In this scenario, nodes intermittently generate CoAP messages to the sink node. In the IoT, such traffic can be seen when information is generated from sensor readings and that information should be exchanged in a database for storing and handling the information. This situation helps to know how the three CC mechanisms perform for various measures of offered traffic loads. In this paper, Confirmable Messages (CON) with end-to-end reliability is used for the Acknowledgement (ACK) and CoAP actuator/toggle URI, which is used to change the state of a variable of the sensor nodes used. All CoAP messages are 71 bytes. Four types of network topologies are used for the analysis of network performance (a) chain with 17 nodes, (b) dumbbell with 21 nodes, (c) grid with 36 nodes (6 X 6), and (d) grid with 49 nodes (7 x 7).

Figure 3 demonstrates the four topologies with one RPL Border Router (Green node), Erbium CoAP client nodes (yellow nodes), primary sink nodes (hubs in blue), and secondary sink nodes for notifications (nodes in blue with a circle). Due to space constraints, grid 6 x 6 topology is shown as a part of the grid 7 x 7 topology, be that as it may, in the simulations, these topologies are separated scenarios.

RPL border router is utilized for starting the Destination Oriented Directed Acyclic Graph (DODAG) and storing the gathered routing data for all nodes that are a part of the DODAG (Winter et al., 2012). In the simulations, the RPL

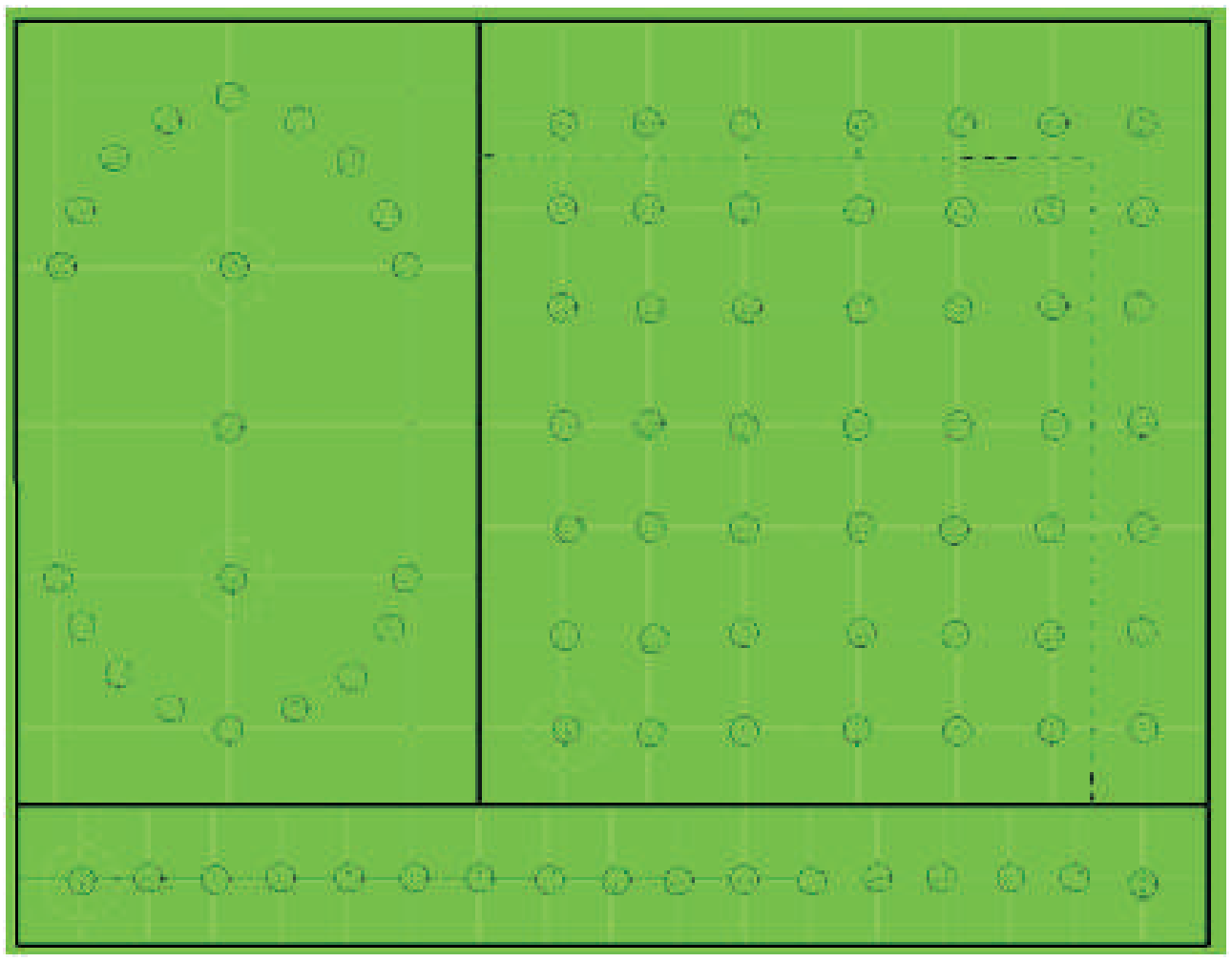

Figure 3. Left: Dumbbell Topology, Right: 7 x 7 Grid and 6 x 6 Grid, and Bottom: Chain Topology. The 6 x 6 Grid is shown as a Subset of the 7 x 7 Grid (The Edges of the Unit Squares are 10 m Long)

border router is characterized to serve just as a relay for CoAP messages, which does not make CoAP messages and is not the destination endpoint of CoAP messages for passing DODAG information to all the CoAP nodes. RPL Node is set amidst all the system topologies.

For investigation of CoAP CC mechanisms performed in this paper, destination endpoint must have a solitary IPV6 address, which in the simulations is a solitary IPv6 sensor node. The proposal of the CoAP base detail to set the NSTART to 1 is set (Osterlind et al., 2006), implying that just a single trade for every destination endpoint is permitted. For radio transmissions, Cooja Simulator for default Unit Disk Graph Medium (UDGM) radio model with the circular transmission is chosen and interference areas are applied. The transmission ranges of the nodes are set to 10 m, which gives a unit square edge in the grids shown in Figure 3. The interference range of the nodes is set to 20 m. Nodes in transmission range can receive the packets and the nodes in interference range cause congestion.

In this paper, simulations are done on Default CoAP, CoCoA, and CoCoA+ CC mechanisms by evaluating in different Link Delivery Ratios (LDR) by 100% LDR, 50% LDR, and 25% LDR. LDR settings are accordingly in transmission success ratio and Reception success ratio in UDGM plugin of Cooja Simulator. When the simulation starts the RPL DODAG formation, it takes up to 60 s for network formation after that CoAP messages will be generated.

The simulations of the constant traffic scenario have a duration of 10 min. The simulation runtime will be given in milliseconds in the simulation script editor of Cooja. After the timeout, the simulation will be stopped and the generated data will be collected from Radio Messages plugin of Cooja simulator into log files. For further evaluation of generated the data, Wireshark Protocol Analyser is used.

The simulations are repeated four times for each configuration of CoAP CC mechanism with different random simulator seed to obtain meaningful average results.

### *3.5 Test Run Configuration*

When a test run is begun, the RPL-border router will make RPL-DODAG by spreading DAG information Objects (DIOs) everywhere throughout the network. The CoAP CC execution will begin after entire formation of RPL-DAG. After that nodes will begin creating CoAP requests. The generation of traffic is execution by running cyclic timers.

These cyclic timers create new CoAP requests upon their depletion. A working of simulation can be seen in Figure 4, where as the network window can observe the RPL-DODAG formation.

For evaluation, the data can be collected from "Radio messages" by applying a 6lowpan filter, and the data will be generated in pcap format, which are stored in the build directory of Contiki. After collecting all the data from each configuration, the generated data will be analyzed in the Wireshark protocol analyzer.

The generated RPL-DODAG and CoAP CON and ACK messages can be observed in Wireshark Protocol Analyser. ICMPV6 means Internet Control Message protocol over IPv6. It is a built-in part of the IPv6 that performs operations like error reporting and other diagnostic functions.

All CoAP requests are POST messages that change the state of sensor nodes. A message ID and node ID of the source node is included in the payload of the CoAP request. Including all headers and the payload of a CoAP request has a size of 71 bytes. Figure 5 shows the data that is generated from one of the simulation tests run with ICMPV6 and CoAP messages of CON and ACK messages.

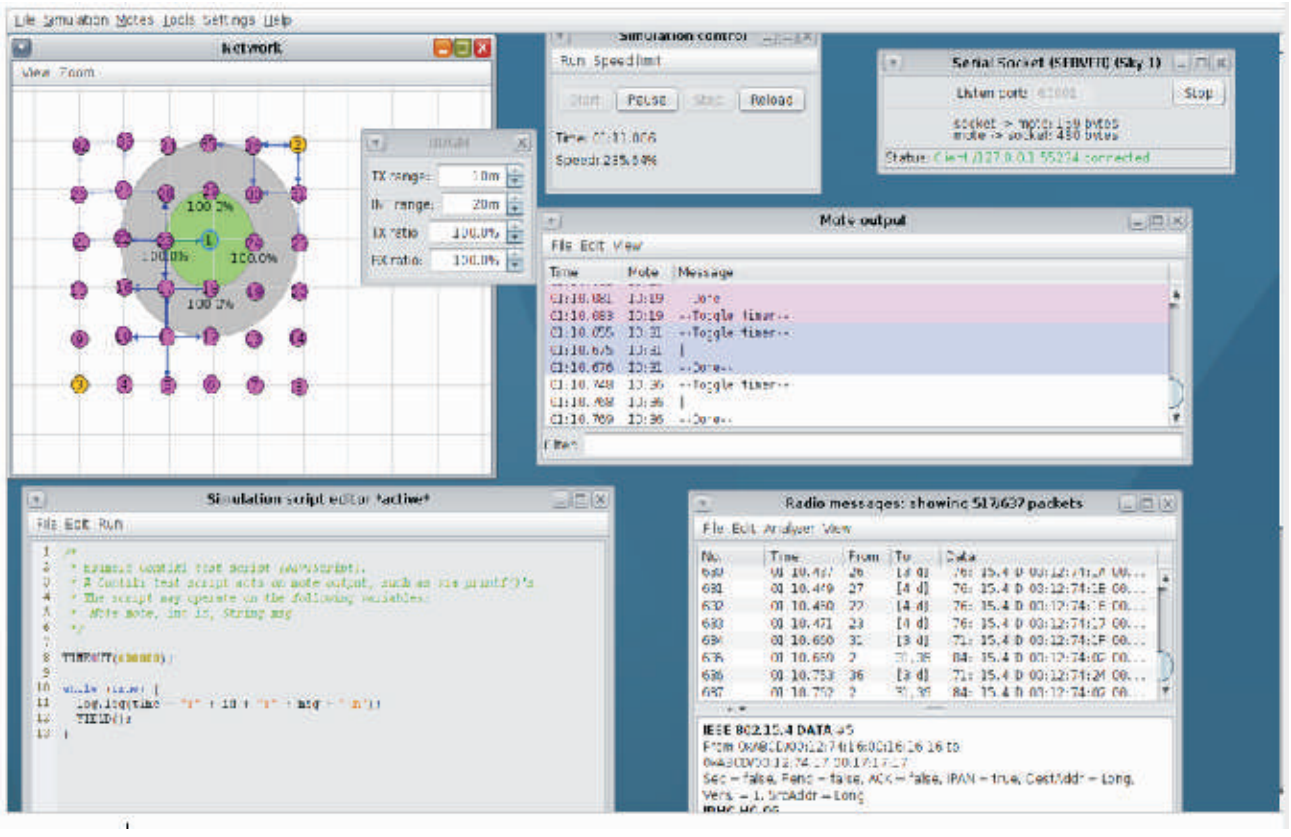

Figure 4. Running Cooja Simulation of Grid 6 x 6 with 100% LDR

In this paper, the average throughput of CoAP CC mechanisms can be evaluated by carried load against the offered load. And the values are collected in average bytes/s. In Wireshark, by using filters the generated traffic loads can be collected as shown in Figure 6 and Figure 7.

For throughput graphs, these traffic loads are varied from 1 kbps to 10 kbps of both offered and carried loads and average values are plotted. These throughput graphs are used to compare network performances of Default CoAP, CoCoA, and CoCoA+.

## 4. Simulation Results

In this section, the results of the evaluation of Default CoAP, CoCoA and CoCoA+ CC mechanisms in reliable Erbium CoAP communications are presented and the results are represented in a graphical format for comparing performances of the CoAP CC mechanisms.

### *4.1 Performance Evaluation*

In the constant traffic scenario, the CC mechanisms are evaluated in four distinctive network topologies. For performance evaluation, the connection of carried load against the offered load is chosen. The offered load alludes to the aggregate sum of data made by the nodes per second. The carried load refers to the aggregate amount of data that is effectively delivered to the sink nodes per second and given in kilobits per second (Betzler et al., 2013).

For analyzing the data created by the nodes, Wireshark protocol analyzer is used for collecting node readings and then the CoAP nodes data transmission is analyzed by getting packet lengths and average bytes per second in the statistics option. Also, to varying traffic loads, the traffic rate at the nodes is changed in accordance with the average network-wide traffic load from 1 kbps to 10 kbps.

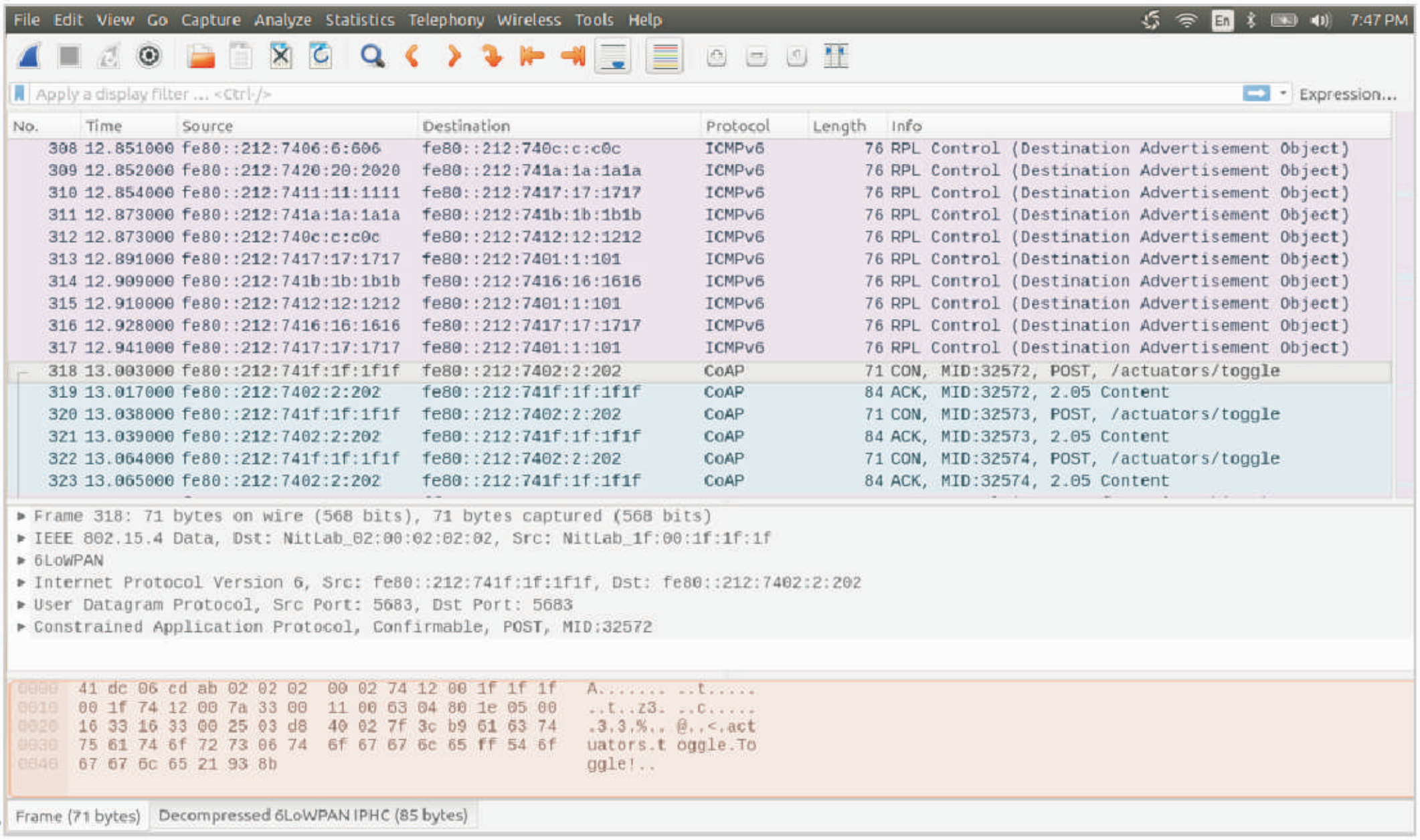

Figure 5. Wireshark Protocol Analyser

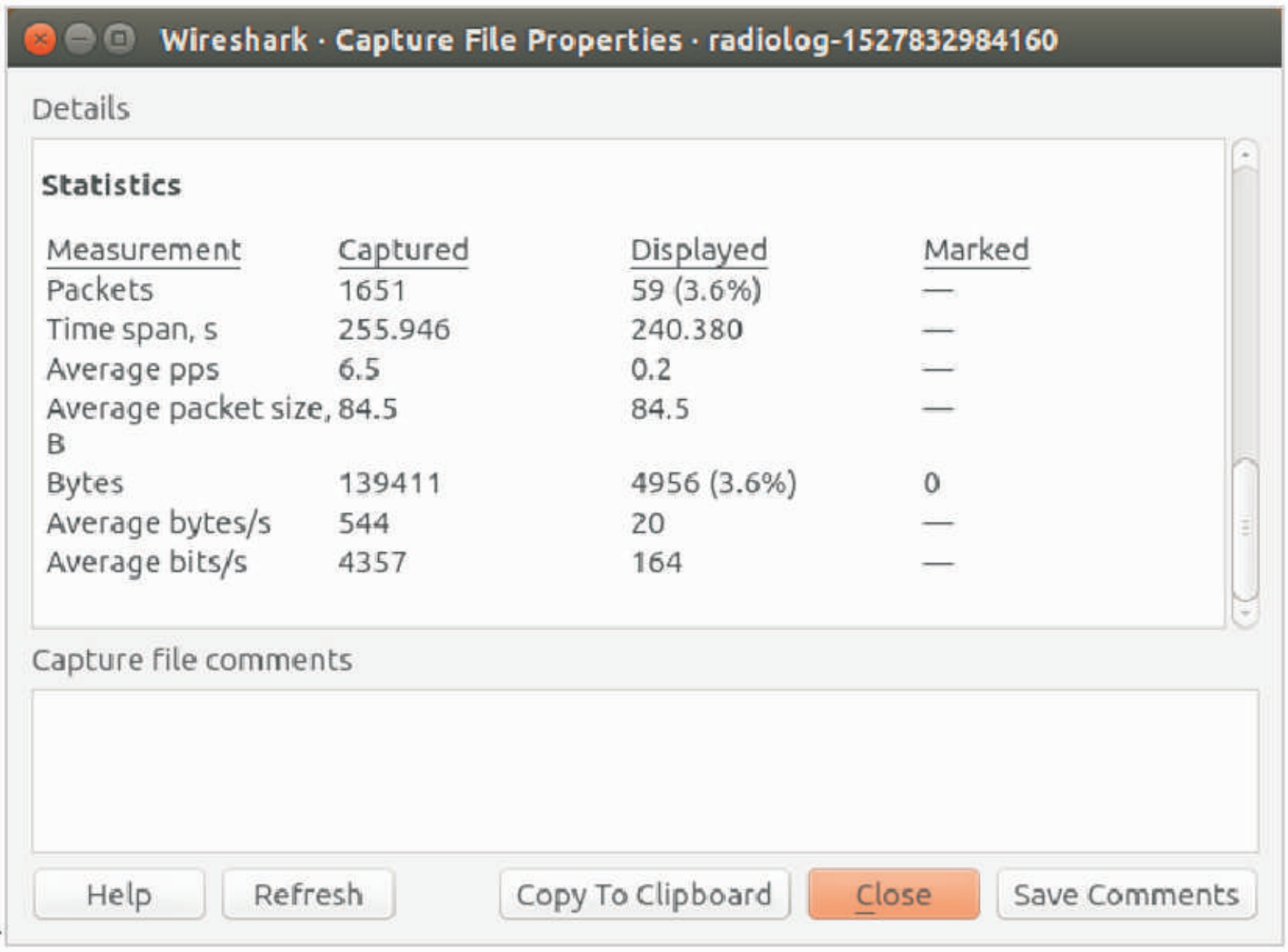

Figure 6. Carried Load

### *4.2 Grid 6 x 6 Network Topology*

Figure 8 illustrates that how carried load advances as the offered load increases when utilizing the three CoAP congestion control algorithms for the Grid 6 x 6 topology evaluated. Since at low traffic load, the network does not achieve a network congestion and all message transmissions across the network are effective, where the carried load will be indistinguishable to the offered load.

As the offered load increases, the proportion of effectively delivered CoAP requests decreases on account of congestion. This is the place that contrasts between the performances of CoAP congestion control mechanisms assessed.

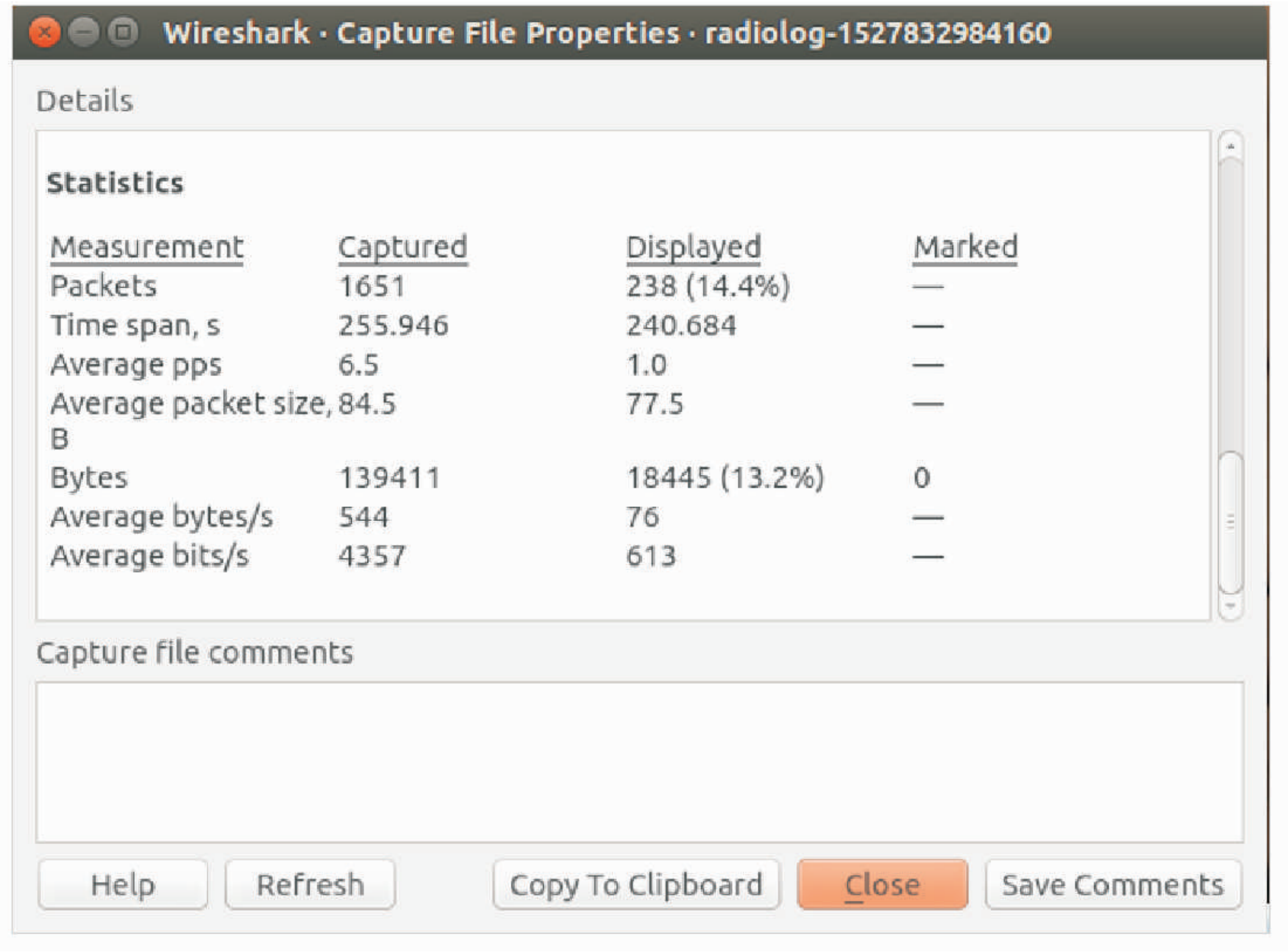

Figure 7. Offered Load

The performances of CC mechanisms CoCoA and CoCoA+ are indistinguishable up to 6 kbps. For offered load up to 2 kbps, the Default CoAP demonstrates a marginally most exceeding terrible execution as far as throughput is concerned. Again the throughput performance reduced to 4 kbps. The quantity of hops among client and sink nodes is little, which is the reason that there is no congestion in form of packet collisions or packet drops, so CoCoA+ mechanism have small RTO estimations.

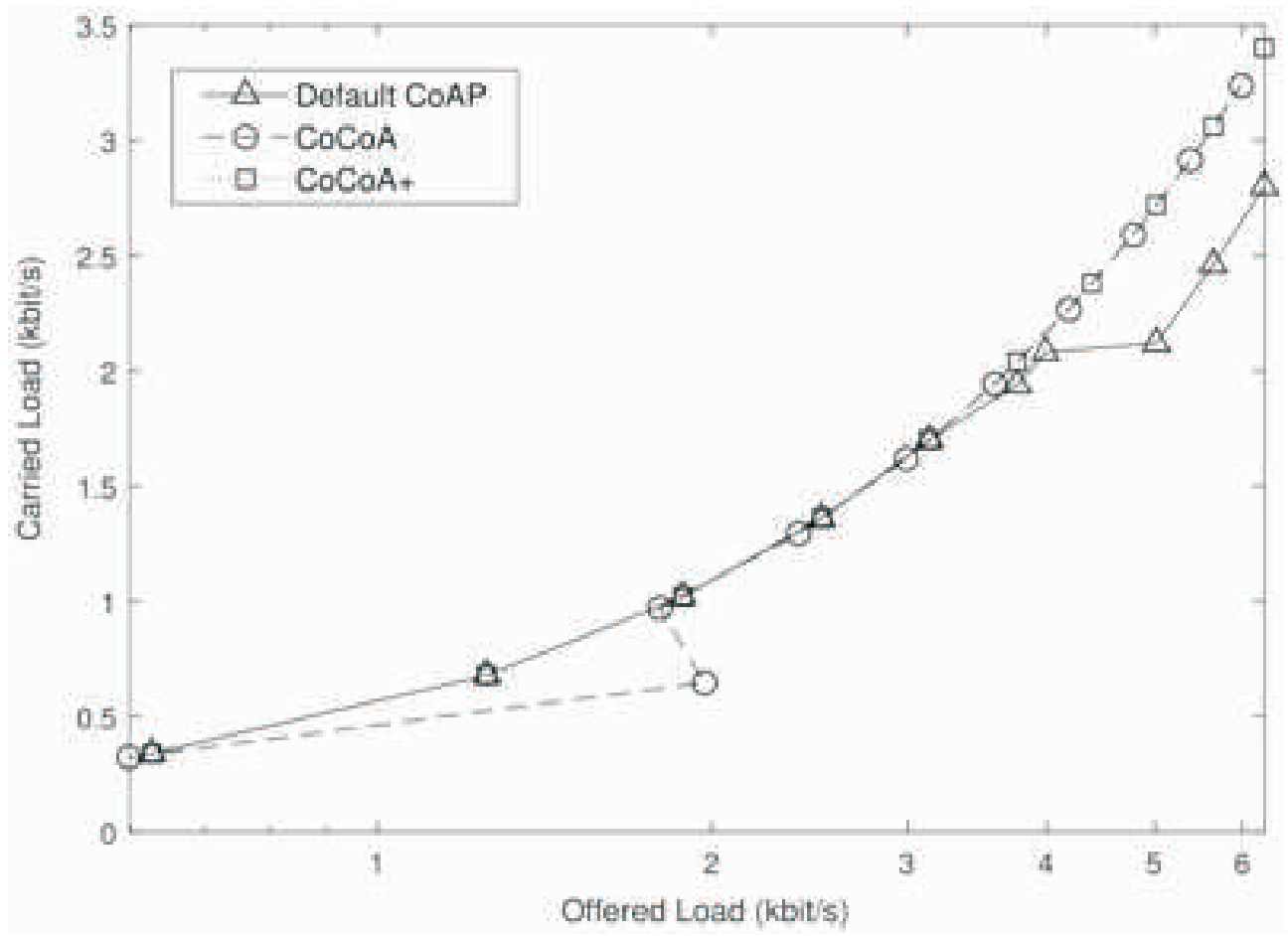

Figure 8. Average Throughput achieved in Grid 6 x 6 Topology with 100% LDR Links

At the point when the assessed RTO gets smaller and achieves the real RTT, the probability for counterfeit retransmission increases and causes the nodes to retransmit a packet they expected to have been lost. These counterfeit retransmissions can prompt dropping in the throughput execution. Packet drops happen when

- The MAC layer buffer overflows
- When CoAP declines to send a packet

Due to the number of parallel transactions allowed for a destination, i.e., NSTART=1, MAC layer buffer overflows are a sign of congestion.

In case of 50% LDR in grid 6 x 6 topology, CoCoA+ outperforms CoCoA and Default CoAP in terms of throughput performance as shown in Figure 9.

CoCoA is able to match the performance of CoCoA+ up to 2 kbps after that it slightly underperforms CoCoA+, but achieves the best performance than Default CoAP. Default CoAP underperforms both CoCoA and CoCoA+ up to 3 kbps offered load after that it achieves performance improvement up to 4 kbps and matches the CoCoA throughput.

In case of grid 6 x 6 topology with 25% LDR as illustrated in Figure 10, Default CoAP, CoCoA, and CoCoA+ performs well up to 1 kbps offered load and among them, CoCoA+ outperforms CoCoA, and Default CoAP up to 3 kbps offered load. CoCoA matches the CoCoA+ performance, but with lower throughput. In Default CoAP it performs equally to CoCoA up to 1 kbps offered load and outperforms CoCoA at 2 kbps and tries to match CoCoA+ performances.

*4.2.1 Chain Network Topology*

In case of Chain topology with 100% LDR as shown in Figure 11, CoCoA+ network performance works better than CoCoA and Default CoAP up to 7 kbps, but in this scenario, Default CoAP matches the CoCoA and CoCoA+ network performance up to 2 kbps and later at 5 kbps offered load it achieves highest throughput value.

Figure 12 illustrates that CoCoA+ underperforms CoCoA up to 2 kbps offered load and CoCoA works similar to CoCoA+ from 2 kbps to 6 kbps offered load. The Default CoAP matches offered load of CoCoA at 3 kbps and unsteadily works up to 6 kbps.

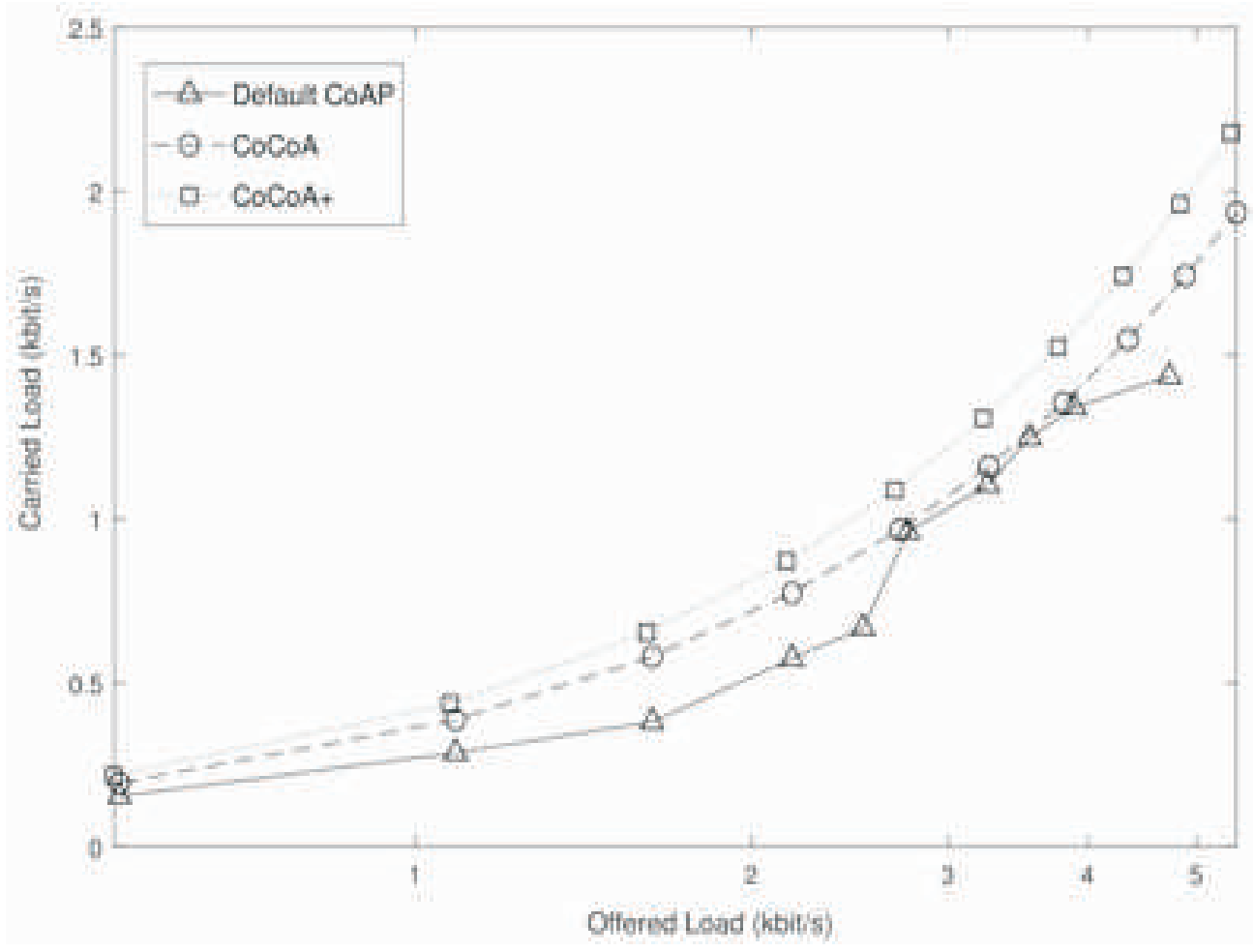

Figure 9. Average Throughput achieved in Grid 6 x 6 Topology with 50% LDR Links

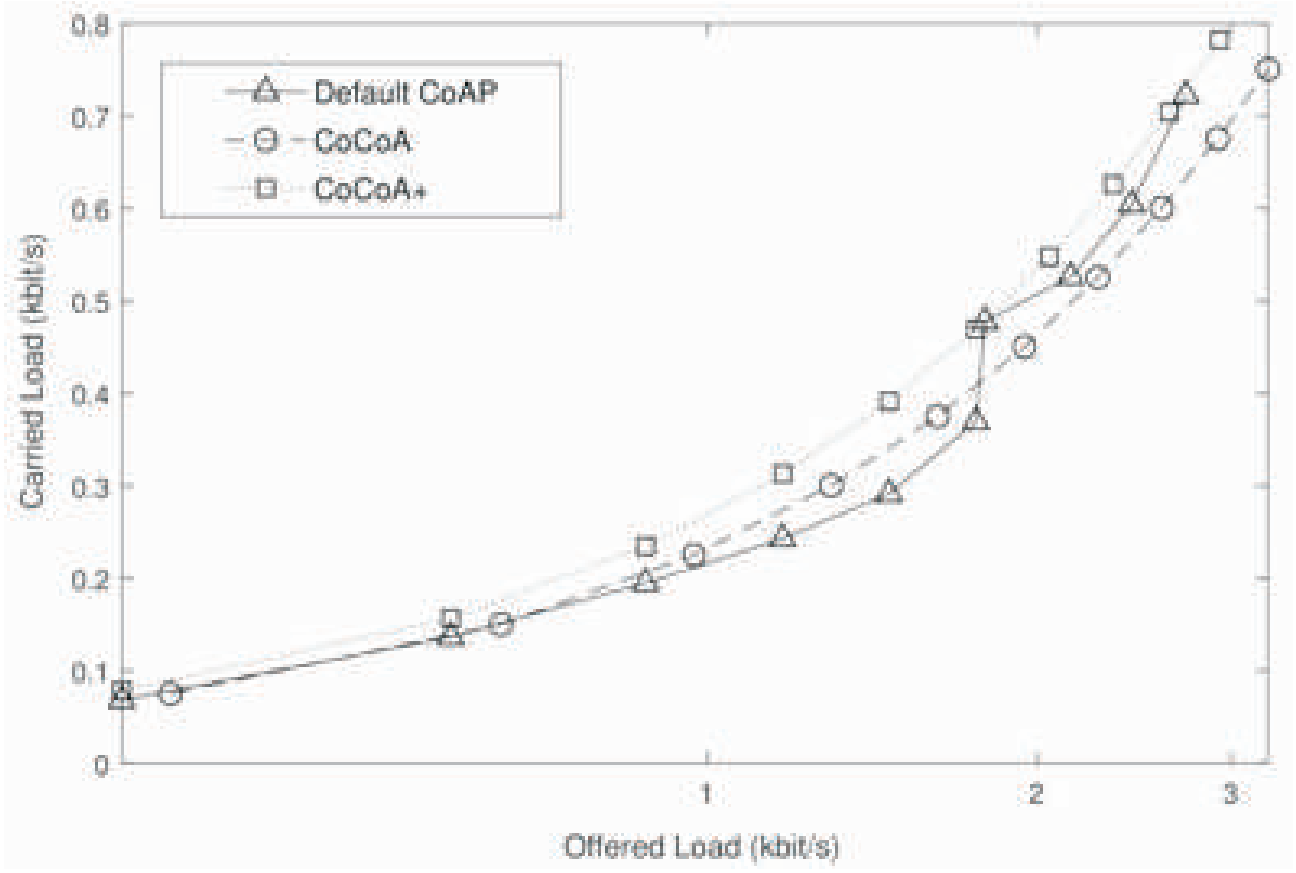

Figure 10. Average Throughput achieved in Grid 6 x 6 Topology with 25% LDR Links

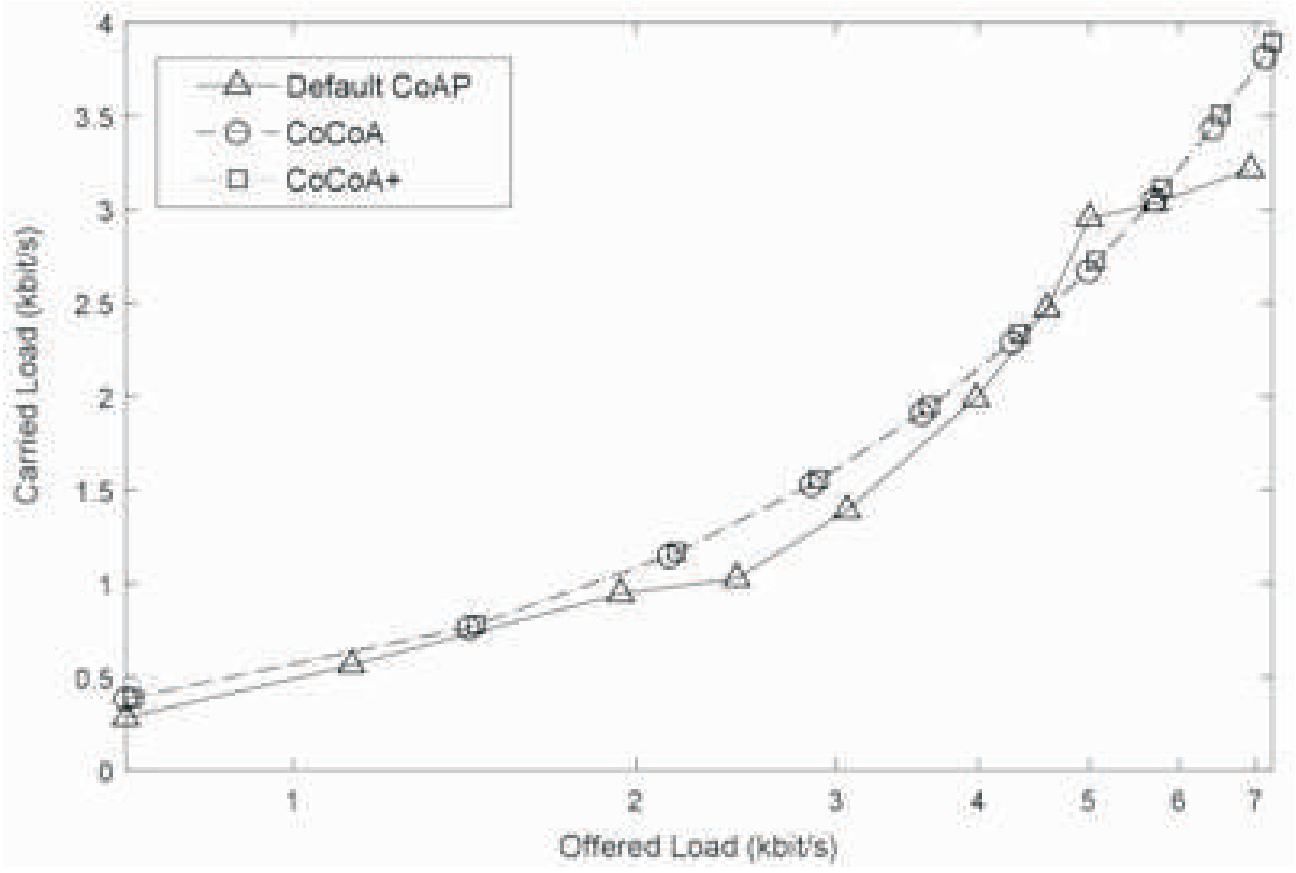

Figure 11. Average Throughput achieved in Chain Topology with 100% LDR Links

In case of Chain topology with 25% LDR as shown in Figure 13, CoCoA+ network performance works better than CoCoA and Default CoAP up to 4 kbps offered load and CoCoA work similar to CoCoA+, but with lower throughput. Default CoAP works with same network performance matching CoCoA up to 2 kbps and unsteadily crosses CoCoA performance at 3 kbps, but underperforms CoCoA+.

*4.2.2 Dumbbell Network Topology*

In the case of the dumbbell topology with 100% LDR, illustrated in Figure 14, nodes on one side of the dumbbell generate data packets for the primary sink on the other side of the dumbbell. In this scenario, The Default CoAP outperforms CoCoA and CoCoA+ up to 2 kbps offered load later underperforms up to 3 kbps and CoCoA+.

Figure 15 illustrates that Default CoAP achieves the highest throughput by outperforming CoCoA and CoCoA+ up to 3 kbps and later its performance got decreased. In this scenario, the CoCoA and CoCoA+ overlaps network performances achieving steady results, but underperforms Default CoAP up to 2 kbps.

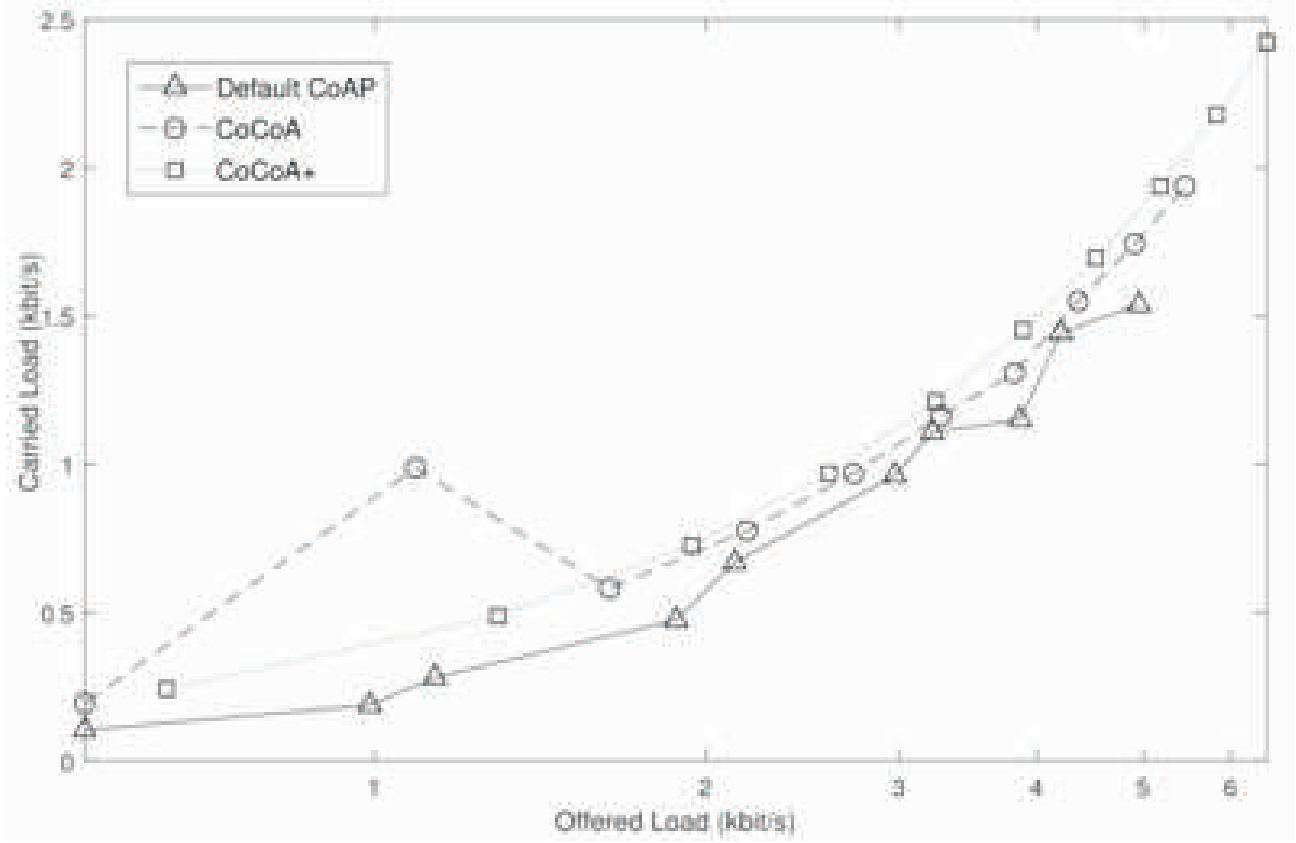

Figure 12. Average Throughput achieved in Chain Topology with 50% LDR Links

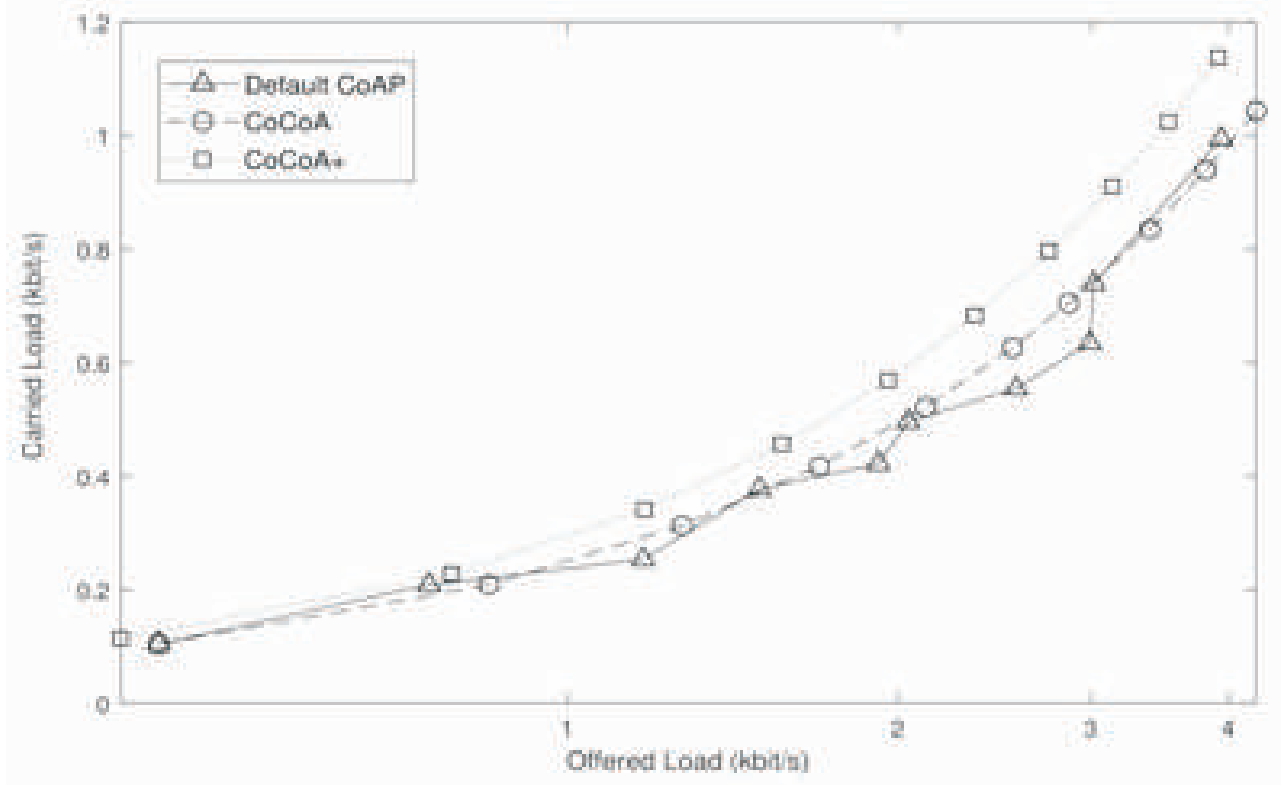

Figure 13. Average Throughput achieved in Chain Topology with 25% LDR Links

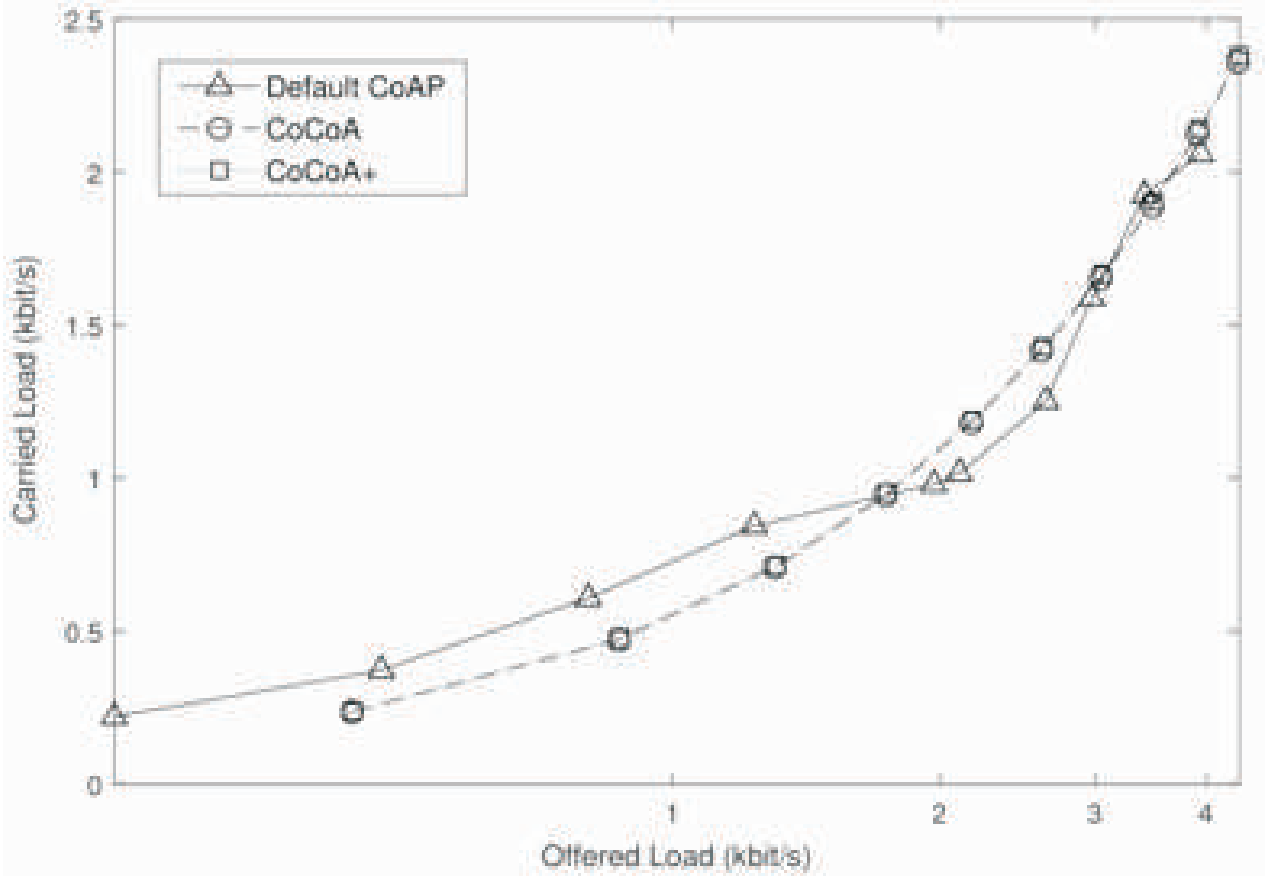

Figure 14. Average Throughput achieved in Dumbbell Topology with 100% LDR Links

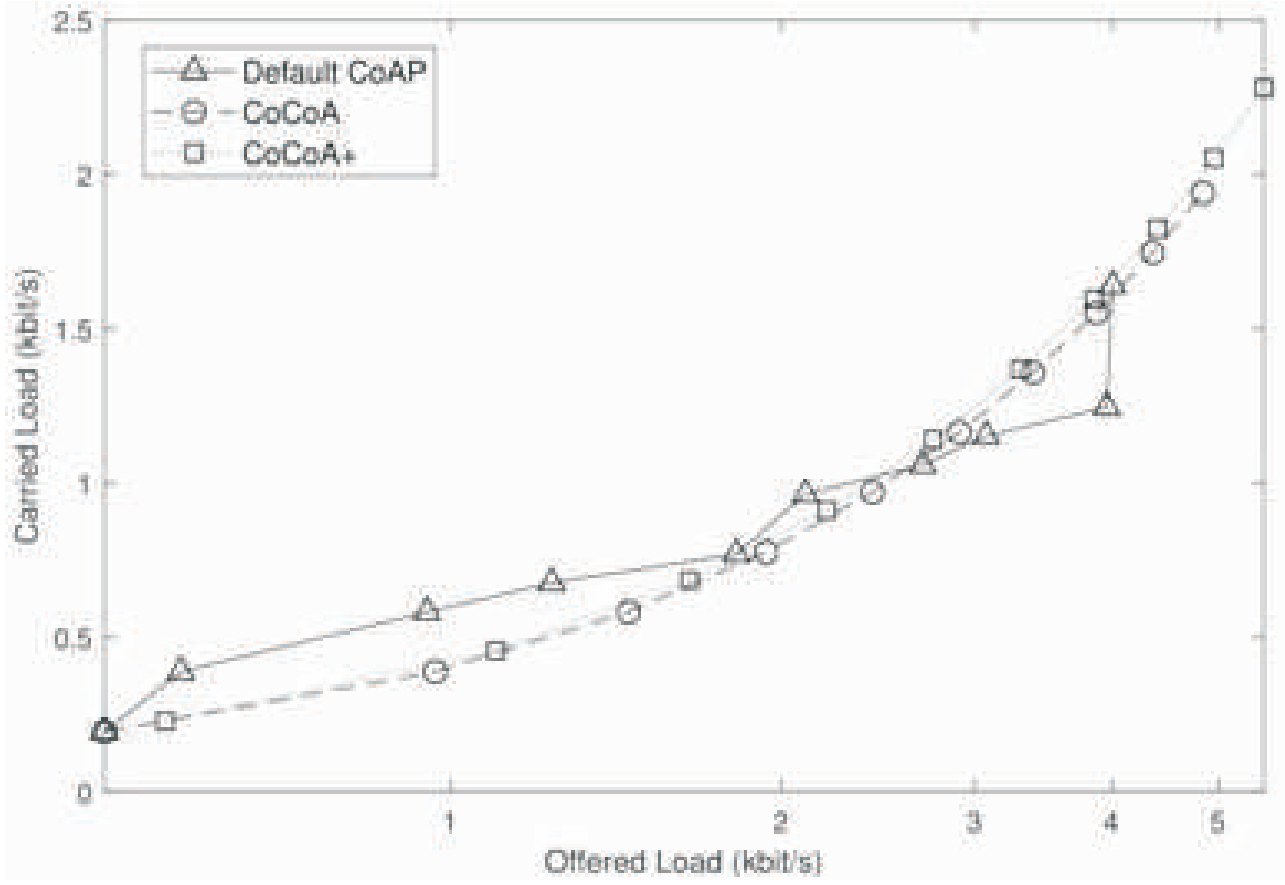

Figure 15. Average Throughput achieved in Dumbbell Topology with 50% LDR Links

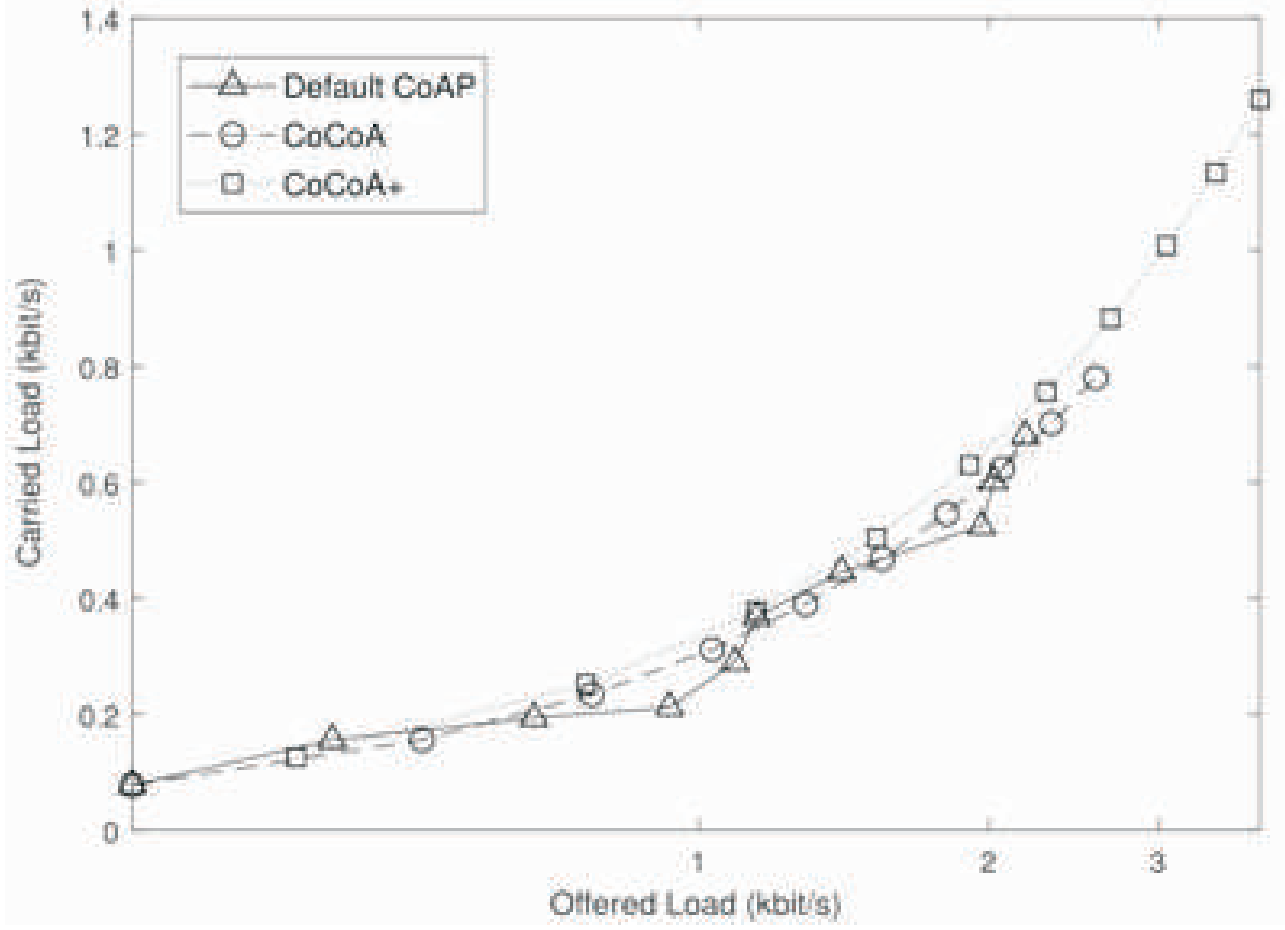

Figure 16. Average Throughput achieved in Dumbbell Topology with 25% LDR Links

Figure 16 illustrates that CoCoA+ achieves highest offered load than Default CoAP and CoCoA. In this scenario, Default CoAP works better than CoCoA.

*4.2.3 Grid 7x7 Network Topology*

Figure 17 illustrates that Default CoAP works better than CoCoA and CoCoA+ up to 2 kbps offered load later underperforms up to 4 kbps. CoCoA network throughput matches the network performance of CoCoA+up to 4 kbps. CoCoA+ performs better in this scenario with steady network throughput although it underperforms Default CoAP up to 2 kbps.

From Figure 18, the CoCoA+ works better than CoCoA and Default CoAP and Default CoAP is the worst performance in this scenario. All CC mechanisms works equal up to 1kbps. CoCoA+ steadily raises its network performance throughout the scenario. At 2 kbps Default CoAP tries to attain performance to CoCoA up to 3 kbps, but fails to catch the CoCoA performance.

CoCoA+ outperforms both CoCoA and Default CoAP up to 2 kbps offered load and CoCoA and default CoAP network performances matches up to 1kbps offered load later underperforms than CoCoA as shown in Figure 19.

The performance of CoAP CC mechanisms, such as Default CoAP, CoCoA, and CoCoA+ in constant traffic scenario and four topologies with three LDRs have been assessed. The outcomes demonstrate that there is no CC mechanism that dependably performs best in all LDR Settings. While CoCoA performs well in an assortment of situations, it frequently performs more awful than default CoAP and a new proposal for CoAP CC mechanism, CoCoA+ performs better than default CoAP in the majority of situations. The results in this paper are an essential and vital advance towards the meaning of a strong alternative to the CC mechanism provided by default CoAP.

## Conclusion

In this paper, a performance investigation is done on Default CoAP, CoCoA, and CoCoA+. By evaluating the generated results of the three CC mechanisms, CoCoA+ can beat Default CoAP and CoCoA in the vast majority of the assessed topologies, constant Traffic scenario, and different Link Delivery Ratio settings. CoCoA+ is flexible against sudden changes in network traffic and adjusts rapidly to various conditions of network congestion. In some scenarios, CoCoA+ does not achieve the same performance as CoCoA, however, it performs still better or very similar to default CoAP.

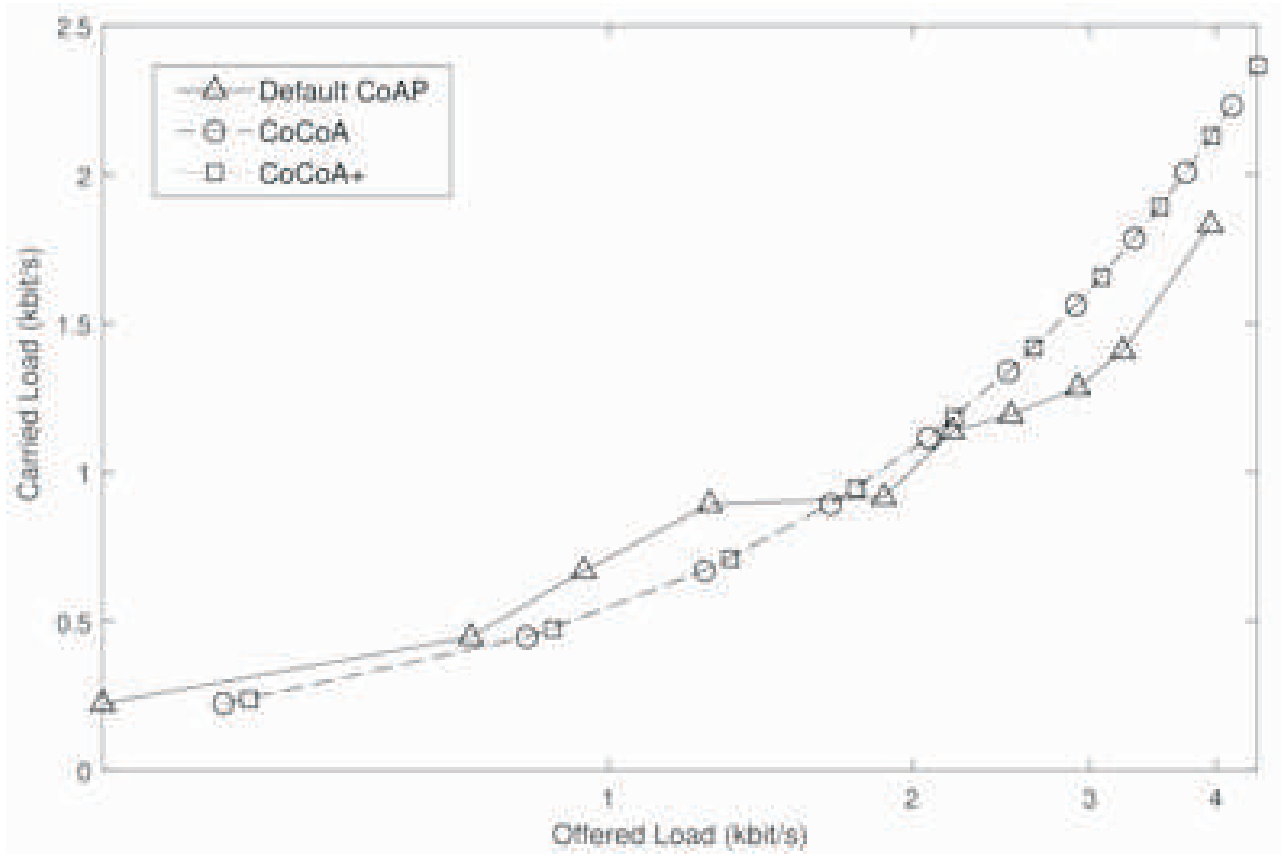

Figure 17. Average Throughput Achieved in Grid 7 x 7 Topology with 100% LDR Links

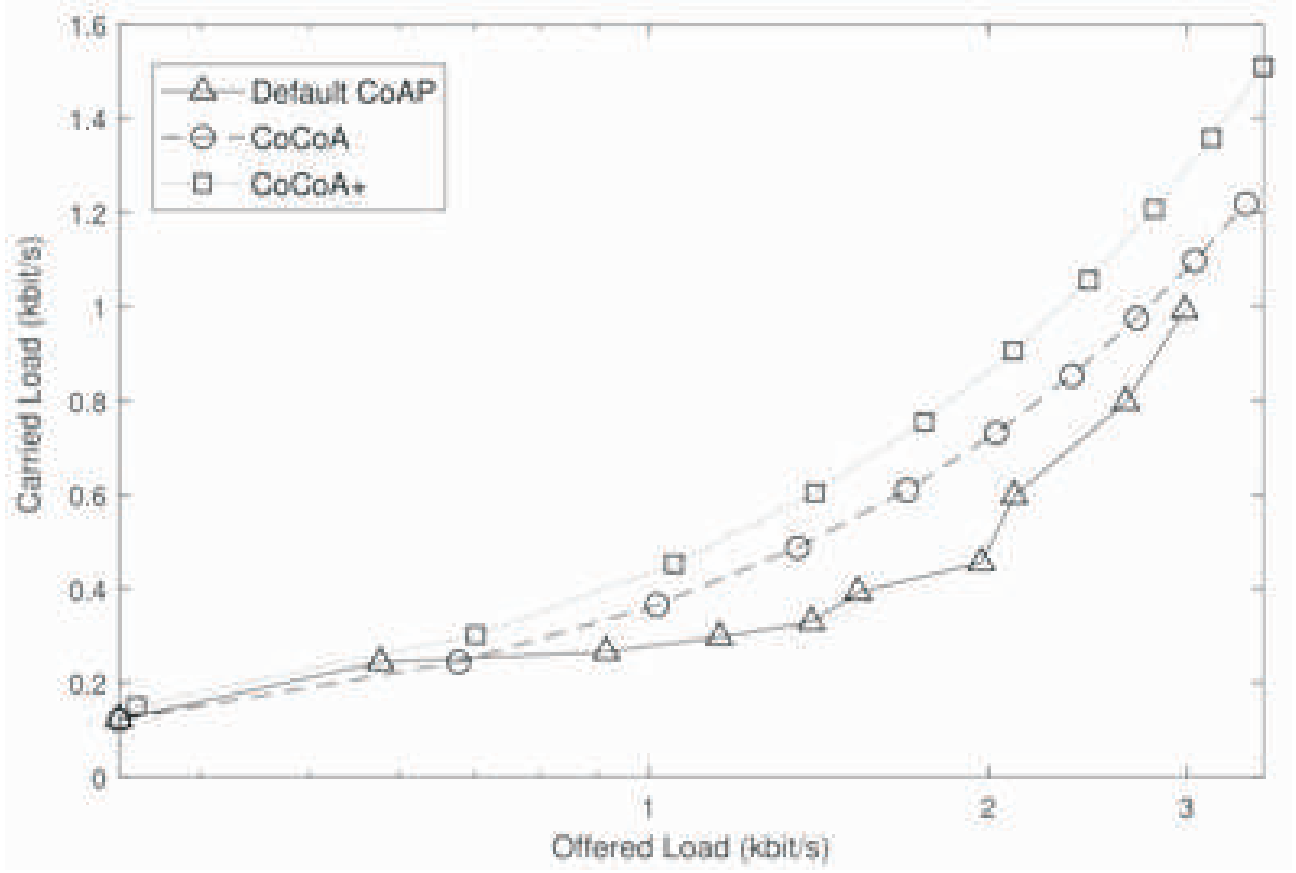

Figure 18. Average Throughput achieved in Grid 7 x 7 Topology with 50% LDR Links

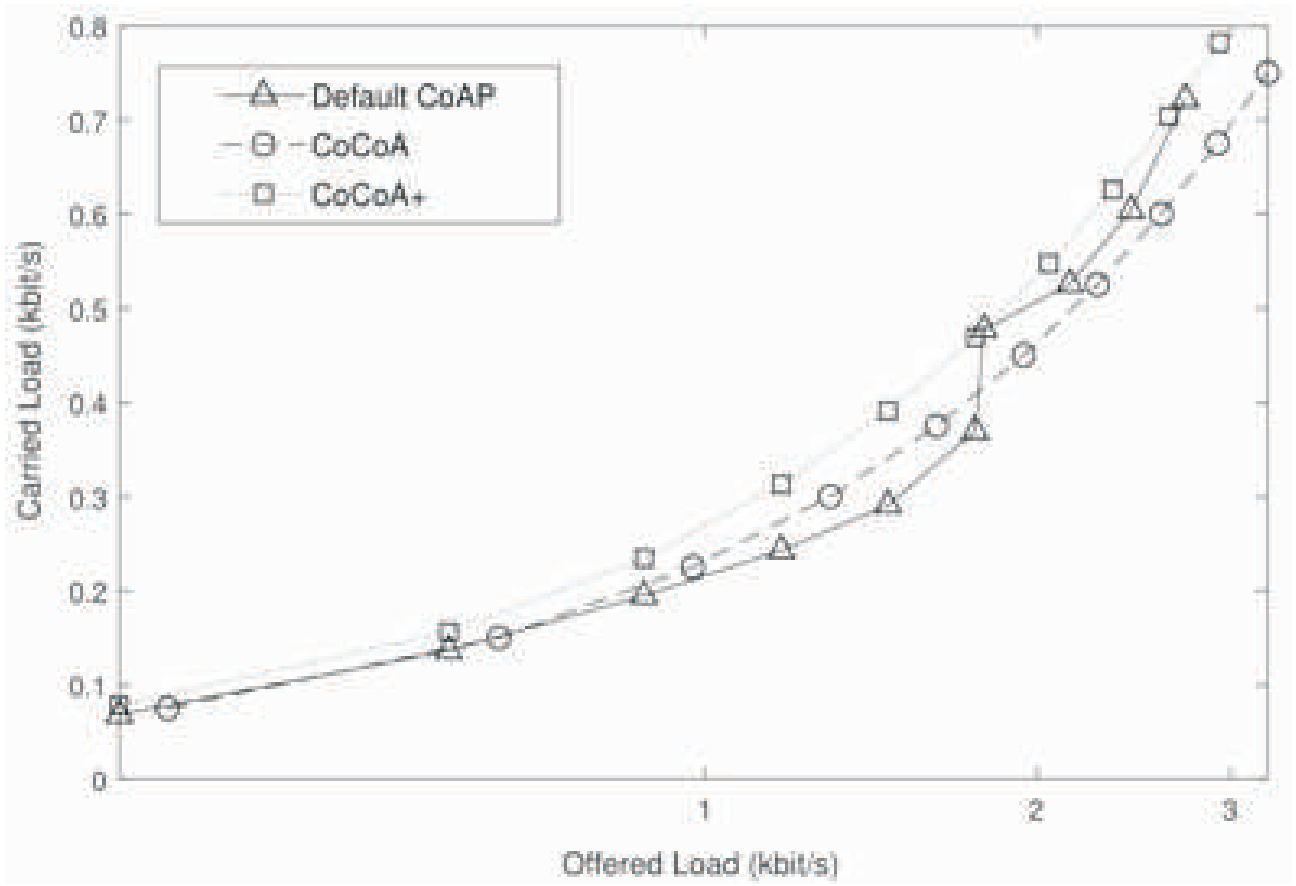

Figure 19. Average Throughput achieved in Grid 7 x 7 Topology with 25% LDR Links

Considering that the main criterion is to propose a suitable congestion control mechanism for CoAP, CoCoA is not an appropriate mechanism because it is unable to provide consistently better performance than Default CoAP. In light of this paradigm and the outcomes acquired in this paper, CoCoA+ is considered to be an exceptionally strong and promising proposition for an advanced CC mechanism for CoAP.

### *Future Work*

As future work, the CoAP advanced congestion control mechanisms have to be evaluated in more network topologies and need to be evaluated in real IoT constrained devices for finding the interactions over the internet will be important to affirm the performances of CC mechanisms. Moreover, investigations of newly proposed CoCoA+ parameters are required.

## ABOUT THE AUTHORS

*Chandra Sekhar Sanaboina is currently working as an Assistant Professor in the Department of Computer science and Engineering at JNTUK, Kakinada, Andhra Pradesh, India. He obtained his B.Tech. in Electronics and Computer Science Engineering in 2005 and M.Tech in Computer Science and Engineering from Vellore Institute of Technology in 2008. He has over 10 years of teaching experience and his areas of interests include Wireless Sensor Networks, Internet of Things, Machine Learning, and Artificial Intelligence.*

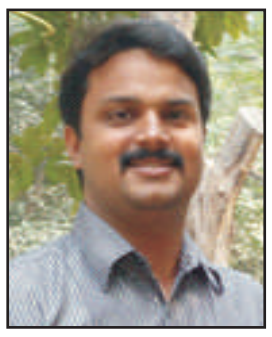

*Tejeswar Eluri is currently pursuing M.Tech in Computer Science and Engineering from UECK, JNTUK, Kakinada, Andhra Pradesh, India. He has completed his B.Tech in Electronics and Communications Engineering from Chaitanya Institute of Science & Technology, Kakinada. He has completed M.Tech Project under the guidance of Chandra Sekhar Sanaboina, Assistant Professor in the Department of Computer Science and Engineering, JNTUK, Kakinada. His areas of interests include Wireless Sensor Networks, Internet of Things, and Image Processing.*

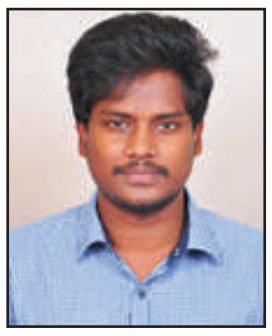